# Room-temperature skyrmion phase in bulk $Cu_2OSeO_3$ under high pressures

Short title: Room-temperature skyrmion phase under pressures

Liangzi Deng,[1] Hung-Cheng Wu,[1,2] Alexander P. Litvinchuk,[1] Noah F. Q. Yuan,[3] Jey-Jau Lee,[4] Rabin Dahal,[1] Helmuth Berger,[5] Hung-Duen Yang,[2,6] Ching-Wu Chu[1,7,*]

[1]Texas Center for Superconductivity and Department of Physics, University of Houston, Houston, Texas 77204-5002, USA
[2]Department of Physics, National Sun Yat-sen University, Kaohsiung, 80424, Taiwan
[3]Department of Physics, Massachusetts Institute of Technology, Cambridge, Massachusetts 02139, USA
[4]National Synchrotron Radiation Research Center, Hsinchu 30076, Taiwan
[5]Institute of Physics of Complex Matter, Ecole Polytechnique Federal de Lausanne, CH-1015 Lausanne, Switzerland
[6]Center of Crystal Research, National Sun Yat-sen University, Kaohsiung, 80424, Taiwan
[7]Lawrence Berkeley National Laboratory, Berkeley, California 94720, USA

*Corresponding author. Email: cwchu@uh.edu (C.-W.C.)

**Abstract**
A skyrmion state in a non-centrosymmetric helimagnet displays topologically protected spin textures with profound technological implications for high density information storage, ultrafast spintronics, and effective microwave devices. Usually, its equilibrium state in a bulk helimagnet occurs only over a very restricted magnetic-field–temperature phase space and often in the low temperature region near the magnetic transition temperature $T_c$. We have expanded and enhanced the skyrmion phase region from the small range of 55-58.5 K to 5-300 K in single-crystalline $Cu_2OSeO_3$ by pressures up to 42.1 GPa through a series of phase transitions from the cubic $P2_13$, through orthorhombic $P2_12_12_1$ and monoclinic $P2_1$, and finally to the triclinic $P1$ phase, using our newly developed ultrasensitive high-pressure magnetization technique. The results are in agreement with our Ginzburg-Landau free energy analyses, showing that pressures tend to stabilize the skyrmion states and at higher temperatures. The observations also indicate that the skyrmion state can be achieved at higher temperatures in various crystal symmetries, suggesting the insensitivity of skyrmions to the underlying crystal lattices and thus the possible more ubiquitous presence of skyrmions in helimagnets.

**Significance**

Skyrmion materials hold great promise for information technology due to the extremely low current needed to modify the spin configurations and the small size of magnetic domains. To facilitate their application, one great challenge is to break the magnetic-field–temperature phase space restriction for the skyrmion state. We found that the temperature region for the skyrmion phase in bulk $Cu_2OSeO_3$ can be greatly enhanced under physical pressure, making applications more practical by the use of strained heterostructures, for example. The observation of additional structures suggests that the skyrmion state may be insensitive to the underlying crystal structure. This work will stimulate further research on finding skyrmion materials with different crystal structures and retaining the room-temperature skyrmion state at ambient condition.

**Introduction**

In a non-centrosymmetric helimagnetic compound, the complex competitions among the various magnetic interactions in decreasing strengths, i.e. the exchange-interaction, the Dzyaloshinskii-Moriya (DM) spin-orbit interaction, and the crystalline anisotropy, result in a generic but complex magnetic-field (H)–temperature (T) phase diagram. For instance, on cooling to below the magnetic transition temperature $T_c$, $Cu_2OSeO_3$ undergoes a paramagnetic-to-helical magnetic transition in a low H less than ~ 0.5 kOe, but a paramagnetic-to-conical magnetic transition in an intermediate H below ~ 2 kOe, and a paramagnetic-to-ferrimagnetic transition in a large H above ~ 2 kOe (1-4). In this generic H-T phase diagram, the skyrmion phase occurs in a very restricted region near $T_c$ ~ 58 K as depicted schematically in Fig. 1. Magnetic skyrmions on the scale of ~ tens of nm emerge with vortex-like spin textures and form the skyrmion lattice state, which has been detected by means of small-angle neutron scattering (SANS) (2, 5), resonant elastic X-ray scattering (REXS) (6, 7), Lorentz force transmission electron microscopy (LTEM) (1, 8, 9), magnetic force microscopy (MFM) (10, 11), electron holography (12), optical polarization rotation measurements (13), and magnetization measurements (1, 3, 4, 14, 15). As a result, great potential has been envisioned for skyrmions for high-density information storage, ultrafast spintronics, and efficient microwave devices (16, 17). To facilitate such a vision, we have decided to enhance the equilibrium skyrmion phase space in bulk $Cu_2OSeO_3$ to a broader and higher temperature region by the application of pressure. We have successfully increased the skyrmion temperature space of $Cu_2OSeO_3$ from 55-58.5 K to 5-300 K, up to room temperature, under pressures up to 42.1 GPa, following a series of structural phase transitions, in agreement with predictions based on the Ginzburg-Landau free energy consideration. The observations will allow for easier device operations and show that high-temperature skyrmion lattices may be found in more helimagnets with different structures.

**Results and Discussion**

**Magnetic studies at ambient and under high pressure**. We have adopted the magnetic technique to identify the skyrmion phase in single-crystal $Cu_2OSeO_3$ by measuring the isothermal ac magnetic susceptibility $\chi'_{ac}(H)_{T,P}$ with an ac-modulation field of 3 Oe at 10 Hz and dc magnetization $M(H)_{T,P}$ as a function of magnetic field $0 < H \leq 1$ kOe at different temperatures T up to 300 K and pressures P up to 42.1 GPa, respectively. All magnetization measurements have been carried out in the zero-field-cooled (ZFC) mode to eliminate the possible H- and/or T-history dependence of the skyrmion state (18). Representative $\chi'_{ac}(H)_{T,0}$ results of $Cu_2OSeO_3$ at ambient pressure and at different temperatures are shown in Fig. 1A. The magnetic fields ($H_{A1}$, $H_{A2}$, $H_{c1}$, and $H_{c2}$) that define the various phase boundaries at ambient are indicated by the arrows in the same figure. As exhibited in Fig. 1A, $\chi'_{ac}(H)_{T,0}$ displays anomalous behavior within a narrow field region ($H_{A1}$, $H_{A2}$) or a $\Delta H \equiv H_{A1}-H_{A2}$ region ~ 200 Oe, over a small temperature range ($T_{A1}$, $T_{A2}$) or a $\Delta T \equiv T_{A2}-T_{A1} = 58.5-55$ K range ~ 3.5 K, where the skyrmion state has been shown to exist by LTEM (1) and SANS (2). These results are summarized in Fig. 1B, which shows the skyrmion state embedded in the magnetic phase diagram of $Cu_2OSeO_3$ together with the known helical, conical, and ferromagnetic phases, in agreement with previous reports (1-4). The schematics for the conical phase, the skyrmion phase, and the skyrmion core based on our model calculation are displayed in Fig. 1C. The narrow field-temperature window for the skyrmion state is also evident from the isothermal $M(H)_{57 K, 0}$ at 57 K that displays the rapid deepening in the slopes of $M(H)_{57 K,0}$ (Fig. 1D), $dM(H)_{57 K,0}/dH$ (Fig. 1E), and $\chi(H)_{57 K,0}$ (Fig. 1F), all in $\Delta H$ ~ 200 Oe and ~ 0.11 kOe < H < 0.32 kOe. The dip feature in Figs. 1E and 1F highlighted by the blue bars is a signature of the skyrmion phase in $Cu_2OSeO_3$.

Our recently developed ultrasensitive high-pressure magnetization technique (19) was used to investigate the skyrmion phase evolution under high pressure. Under pressure the magnetic phase transition temperature $T_c$, which defines the upper bound of the skyrmion phase, is shifted toward higher temperature as shown in Fig. 2. Above 7.9 GPa, it becomes increasingly difficult to distinguish the magnetic transition from the background signal based on dc magnetization measurements. We have measured the $\chi'_{ac}(H)_{T,P}$ as a function of H at different P up to 42.1 GPa and T up to 300 K and the accompanying H-T regions where the skyrmion state occurs at selected pressures, e.g. 2.5, 7.9, 26.2, and 42.1 GPa, as shown in Figs. 3A-D, respectively. It is clear that the temperature region ($T_{A1}$,$T_{A2}$) for skyrmions, or $\Delta T \equiv (T_{A2}-T_{A1})$, has been expanded from (55 K, 58.5 K), i.e. ~ 3.5 K, at ambient to (5–10 K, > 300 K), i.e. > 290 K, at 42.1 GPa by lowering $T_{A1}$ and raising $T_{A2}$ to above 300 K via pressures. The extension of $T_{A2}$ to above 300 K above 7.9 GPa is in agreement with the results shown in Fig. 2 for $T_c$ and makes the skyrmion state accessible without the aid of liquid cryogen for device applications. At the same time, the field region ($H_{A1}$, $H_{A2}$) or $\Delta H \equiv H_{A2}-H_{A1}$ remains almost unchanged, keeping the accessible field low, in spite of the great expansion in $\Delta T$. The observation is in qualitative agreement with previous reports at low pressures up to 1.4 GPa by Wu et al. (4) and up to 5.7 GPa by Sidorov et al. (20), who raised $T_{A2}$ from ~ 56 K to 60.5 K and $T_c$ to 75 K, respectively. It is worth mentioning that while $T_{A2}$ and $\Delta T$ are observed to increase smoothly with pressure, they suffer a drastic increase at 7.9 GPa, suggesting a possible pressure-induced structure transition to be explored below.

**Synchrotron X-ray measurements under high pressure.** The sudden expansion of $\Delta T$ resulting from the rapid increase of $T_{A2}$ around 7.9 GPa and the precipitous drop of $T_{A1}$ around 26.2 GPa for the skyrmion state strongly suggest possible pressure-induced structure transitions in $Cu_2OSeO_3$. Because of the small volume (~0.003 $mm^3$) of our sample in a high-pressure diamond anvil cell (DAC), we decided to carry out the structural study using synchrotron X-ray diffraction. Room-temperature synchrotron X-ray diffraction with a wavelength of 0.6889 Å (18 keV) was performed and the patterns were analyzed. As shown in Fig. 4A, they display the cubic $P2_13$ phase with the lattice constant a of 8.9193 Å, consistent with a previous report (3). The same crystal structure persists as the pressure increases to 3.96 GPa. However, at 5.28 GPa, new Bragg reflection peaks emerge, indicating the breaking of crystal symmetry. This pattern can be indexed within the orthorhombic phase with the $P2_12_12_1$ space group (losing the 3-fold rotational symmetry) with lattice parameters a = 8.7988 Å, b = 8.7790 Å, and c = 8.7409 Å. At ~ 7.01 GPa, $Cu_2OSeO_3$ undergoes a second structural transition to the monoclinic phase with the $P2_1$ space group (losing the $2_1$ screw axis symmetry). Consequently, the schematic diagram of relevant pressure-induced structural phases was established and is shown in Fig. 4B, where the cubic $P2_13$ phase, the orthorhombic $P2_12_12_1$, and the monoclinic $P2_1$ are marked in black, blue, and red colors, respectively. The results add two additional structure phases below 11 GPa, the limit of our synchrotron XRD experiment, that can host the skyrmions.

**Raman measurements under high pressure.** Raman spectra taken with increasing pressure are presented in Fig. 5A and provide further experimental evidence for the existence of several crystallographic phases of $Cu_2OSeO_3$ in the pressure range under investigation as described above, as well as beyond that allowed by synchrotron XRD experiments. First, we note that a large number of observed modes, as well as their relative intensities, are in very good agreement with those reported earlier (21). Spectra shown in Fig. 5A suggest the existence of the several distinct crystallographic phases as pressure increases, namely $P2_13$-cubic below 5.5 GPa, $P2_1$-monoclinic between 7.3 GPa and 23.1 GPa, and finally $P1$-triclinic above 24.3 GPa. The phase diagrams based on the Raman and synchrotron XRD results are consistent with one another. They are summarized in Fig. 5B and more detailed analyses can be found in SI Appendix, Section S1. Raman spectra were also recorded upon releasing the pressure (SI Appendix, Fig. S1), signaling the existence of a few additional structural transitions. A direct comparison between the initial and final ambient-pressure spectra (SI Appendix, Fig. S2) demonstrates a clear difference, indicating the irreversibility of the loading-unloading process.

The phase transitions show an interesting correlation with the magnetic measurement results. Within Phases I and II, the temperature region ($T_{A1}$, $T_{A2}$) for skyrmions expands slowly and $T_{A2}$ increases as the pressure increases. At the transition to Phase III, $T_{A2}$ suddenly extends to room temperature, the highest temperature measured in this experiment. At the transition to Phase IV, $T_{A1}$ extends to a lower temperature, i.e. 5–10 K, while $T_{A2}$ remains at room temperature.

**Ginzburg-Landau free energy analysis.** The above experimental observations show that the skyrmion phase is stabilized in a much greater H-T phase space by pressure. To get an insight into the possible mechanism of stabilization of the skyrmion state of $Cu_2OSeO_3$ by pressure, we carried out a Ginzburg-Landau free energy analysis. Within each phase, the explicit form of DM interaction was derived (SI Appendix, Section S2), and it was demonstrated that a stronger DM interaction can enlarge the temperature region of skyrmion phase, which may lead to the gradual expansion of the skyrmion temperature range under pressure. More importantly, we also showed that DM interaction can be enhanced dramatically and abruptly when the system goes through a structural transition by breaking crystal symmetry, resulting in the sudden expansion of the skyrmion temperature region near the structural phase transition, which may correspond to the sudden extension of the skyrmion temperature region at the transitions to Phase III and Phase IV.

For a non-centrosymmetric helimagnet, the Ginzburg-Landau free energy can have the form $f = f_0 + J(\nabla \boldsymbol{M})^2 + D\boldsymbol{M} \cdot (\nabla \times \boldsymbol{M})$ (22-24), where $\boldsymbol{M}$ is the magnetization, $(\nabla \boldsymbol{M})^2 \equiv \sum_{ij}(\partial_i M_j)^2$, and $J$ and $D$ denote Heisenberg and DM interactions among magnetic moments, respectively. The uniform part of the free energy is $f_0 = a(T - T_m)M^2 + bM^4$ at temperature $T$, with phenomenological parameters $a$, $b$, and $T_m$ determined by microscopic details of the magnet.

From the free energy described above, another temperature scale $T_0 = \frac{D^2}{Ja}$ can be obtained by dimensional analysis. As demonstrated in SI Appendix, Section S2, the temperature region for skyrmions $\Delta T \equiv (T_{A2}\text{-}T_{A1})$ should be proportional to the temperature scale $T_0$. It should be noted that $D$ measures the breaking of centrosymmetric symmetry, so as we increase the pressure and reduce the symmetry, $D$ should increase and show sudden jumps near structural transitions. As a result, the structural transitions induced by high pressure lead to $\Delta T_I < \Delta T_{II} < \Delta T_{III} < \Delta T_{IV}$, where $\Delta T_i$ represents the temperature region for skyrmions in Phase i (i = I to IV). These results are consistent with experimental results presented in Figs. 3-5. In general, the critical temperatures can be computed by numerical methods such as Monte Carlo simulations (25). It is worth mentioning that since the surface and/or interface can affect DM interaction (1), perhaps associated with the strain relaxation due to the sample thickness, the change of sample thickness under pressure may also play an important role in the enhanced skyrmion region we observed.

**Summary**

In summary, we have carried out a systematic study of the high-pressure effect on the skyrmion phase in $Cu_2OSeO_3$ via magnetization, synchrotron X-ray and Raman spectra measurements. We also performed Ginzburg-Landau free energy analysis on the crystal symmetry and DM interaction terms, which affect the temperature region for the skyrmion phase. We found that pressure favors the skyrmion phase in this system and that the structural transitions induced by higher pressure help to extend the temperature region of the skyrmion phase immensely. SANS measurements under high pressure are currently being taken to obtain direct evidence for the pressure-induced room-temperature skyrmion phase. The results demonstrate the potential role of stress/strain in the stabilization of the skyrmion phase at room temperature. They further suggest that chemical pressure may be deployed to retain the phase at ambient condition, similar to stabilizing high temperature superconductivity by replacing ions in the compound with smaller iso-valent ions (26, 27). The appearance of the skyrmion phase in different crystal structures of $Cu_2OSeO_3$ implies that the skyrmion phase is not sensitive to the symmetry of the underlying host, suggesting that more host materials can be found. New synthesis routes, *e.g.* chemical doping/catalyzing or pressure quenching, may be helpful in retaining the high-pressure and room-temperature skyrmion phase in this and other materials at ambient pressure for applications.

**Materials and Methods**

**Crystal growth.** The single-crystalline samples of $Cu_2OSeO_3$ were grown from the polycrystalline compound by the chemical vapor deposition method. Pure polycrystalline $Cu_2OSeO_3$ was synthesized using the solid-state-reaction technique by heating pressed pellets of a stoichiometric mixture of high-purity CuO and $SeO_2$ in an evacuated quartz tube at 510–600 °C for 72 hours followed by slowly cooling to room temperature. To improve the sample quality, intermediate breaking, grinding, and pressing of the pellet was performed several times. The X-ray diffraction patterns of the samples at ambient determined by a Rigaku D-MAX-BIII diffractometer show the high quality of the samples investigated.

**Magnetization measurements under high pressure.** The skyrmion state has been shown to be easily identified by the ac susceptibility ($\chi'_{ac}$) as a function of dc magnetic field (H) at different temperature (T). To determine the pressure effect on the skyrmion state we have deployed our ultrasensitive high-pressure magnetization technique using the diamond anvil cell (DAC) incorporated within a Quantum Design Magnetic Property Measurement System (MPMS). The technique allows us to measure the dc and ac magnetization of $Cu_2OSeO_3$ single crystals with diagonal ~100 μm and thickness of a few micrometers at a T between 5 to 300 K for an H up to 1 kOe under pressure up to 42.1 GPa. Since the skyrmion state has been shown to be thermal- and field-history-dependent, we chose to measure all isothermal and isobaric magnetizations, $\chi'_{ac}(H)_{T,P}$, following the zero-field-cooled (ZFC) mode. The pressure experienced by the sample inside the DAC is determined by the fluorescence line of ruby powders and the Raman spectrum from the culet of the top diamond anvil. A pair of 300-μm-diameter culet-sized diamond anvils was used. The gaskets were made from nonmagnetic Ni–Cr–Al alloy. Each gasket was pre-indented to ~20–40 μm in thickness, and a hole ~120-μm in diameter was drilled to serve as the sample chamber. A mixture of methanol and ethanol in a ratio of 4:1 was used as the pressure-transmitting medium.

**High-pressure PXRD.** High-pressure powder X-ray diffraction (PXRD) experiments were performed using the DAC technique at the Taiwan Light Source (TLC) 01C2 powder beamline. Beam size was collimated to 150 μm diameter. The DAC was rocked ± 3° during exposure time to obtain smoother data. Pressure was generated by means of three screw-driven ALMAX type DACs (28) equipped with Boehler-Almax (29) diamond anvils (450 μm culet size). The aperture of the DAC is up to 85 degrees for obtaining high-Q diffraction data. Stainless steel gaskets were pre-indented to about 80-100 μm thickness, drilled in the middle of the indentation to obtain a ~ 200 μm hole, and placed between two diamonds to form a pressure sample chamber. The pressure chamber with the sample and a small amount of ruby powder was loaded and sealed with silicone oil as the pressure medium (Alfa Aesar, Polydimethylsiloxane, trimethylsiloxy terminated, M.W. 410 CAS:43669). Pressure was determined using the ruby R1 fluorescence line as a pressure marker by Raman spectrometer iHR550 (Horiba Jobin Yvon).

**Raman spectroscopy under high pressure.** The pressure cell and sample were prepared using a method similar to that described in the section "Magnetization measurements under high pressure". All light scattering measurements were performed in the back-scattering geometry at room temperature using the triple Raman spectrometer T64000 (Horiba Jobin Yvon) equipped with a microscope, a liquid-nitrogen-cooled charge-coupled-device detector, and an $Ar^+$-ion laser ($\lambda_{exc}$ = 488 nm) as the excitation source. Laser excitation power was kept below 1 mW in order to minimize heating of the sample. The spectral resolution is 1.5 $cm^{-1}$.

**Acknowledgments:** We thank B. Haberl, J. Molaison, L. Debeer-Schmitt, C. Do, C. Cruz, and K. Taddei for efforts on the small-angle neutron scattering experiments under high pressure at Oak Ridge National Laboratory. The work performed at the Texas Center for Superconductivity at the University of Houston is supported by the U.S. Air Force Office of Scientific Research Grant FA9550-15-1-0236, the T. L. L. Temple Foundation, the John J. and Rebecca Moores Endowment, and the State of Texas through the Texas Center for Superconductivity at the University of Houston. The work done at the National Sun Yat-sen University is supported by the Ministry of Science and Technology, Taiwan, under Grant No. MOST 106-2112-M-110-013-MY3. The work carried out at the Department of Physics at the Massachusetts Institute of Technology is supported by DOE Office of Basic Energy Sciences, Division of Materials Sciences and Engineering under Award de-sc0010526, and partly supported by the David and Lucile Packard Foundation.

Figures

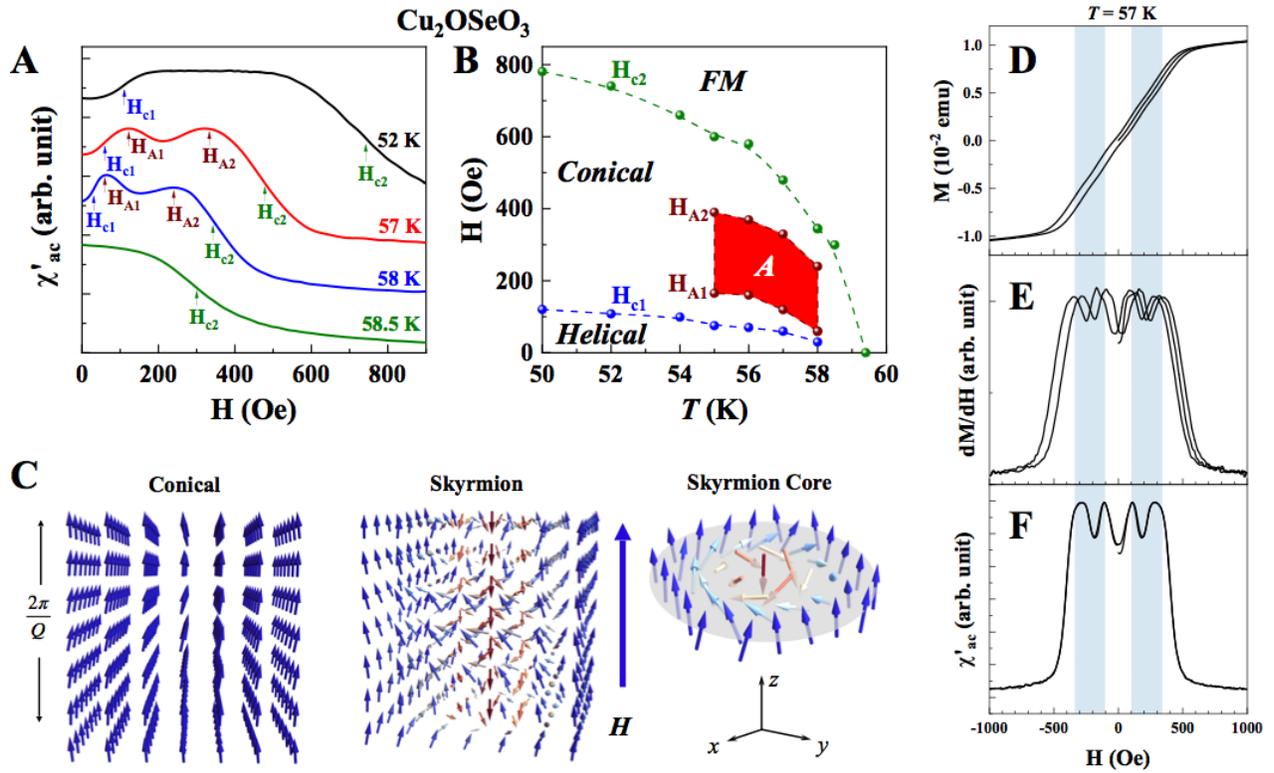

**Fig. 1. Magnetic measurements of $Cu_2OSeO_3$ at ambient pressure**. (**A**) ac susceptibility as a function of dc bias field up to 900 Oe at different temperatures for a single-crystal $Cu_2OSeO_3$ sample at ambient pressure. The curves are shifted vertically within the same scale. (**B**) magnetic phase diagram for a single-crystal $Cu_2OSeO_3$ sample at ambient pressure. (**C**) Schematics for conical and skyrmion phases. The direction of the arrows denotes the direction of magnetization and the color denotes the z-component of magnetization. The conical phase is plotted from the conical ansatz in Section S2, while the skyrmion phase is schematic. (**D**) M, (**E**) dM/dH, and (**F**) $\chi'_{ac}$ as a function of magnetic field up to 1000 Oe at 57 K.

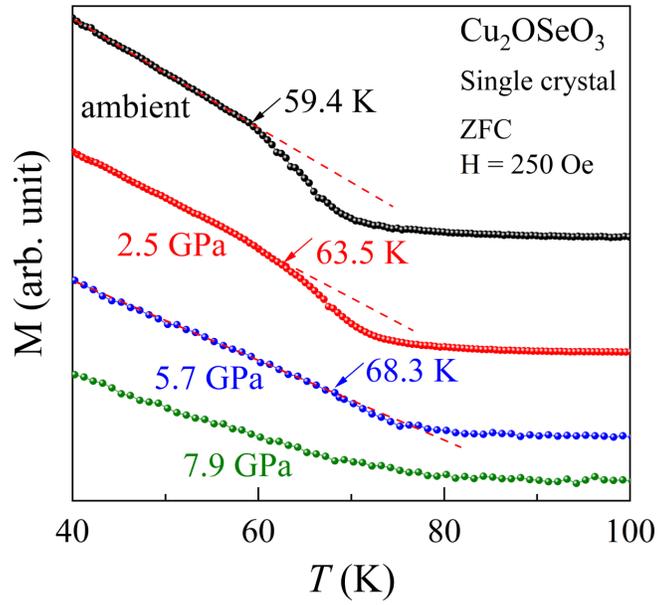

**Fig. 2. DC magnetization measurements of $Cu_2OSeO_3$**. M vs. T measured on a single-crystal sample under magnetic field of 250 Oe and at different pressures up to 7.9 GPa in zero-field-cooled (ZFC) mode. The arrows indicate the ferromagnetic transition temperature at different pressures.

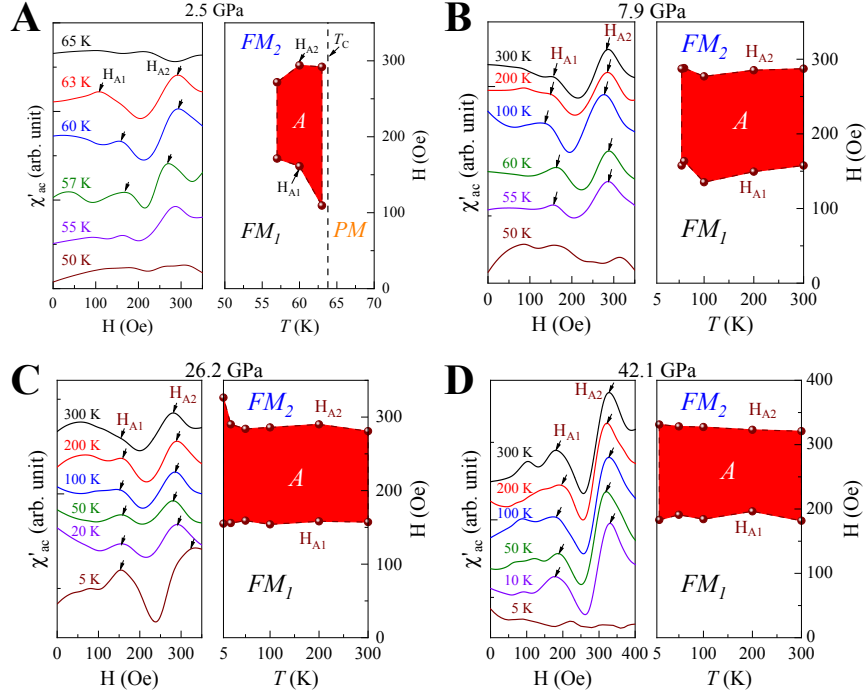

**Fig. 3. ac susceptibility of $Cu_2OSeO_3$ as a function of magnetic field at different critical pressures**. (**A**) 2.5 GPa; (**B**) 7.9 GPa; (**C**) 26.2 GPa; (**D**) 42.1 GPa. The evolution of the "dip figure" indicates that the temperature region for the possible skyrmion state expands under pressure. At 7.9 GPa, the upper limit of the temperature range, $T_{A2}$, increases to 300 K, the highest temperature measured in this experiment. At 26.2 GPa, the lower limit of the temperature range, $T_{A1}$, extends to 5 K. With increasing pressure up to 42.1 GPa, the "dip feature" becomes more pronounced while the temperature range remains at between 5-10 K and 300 K.

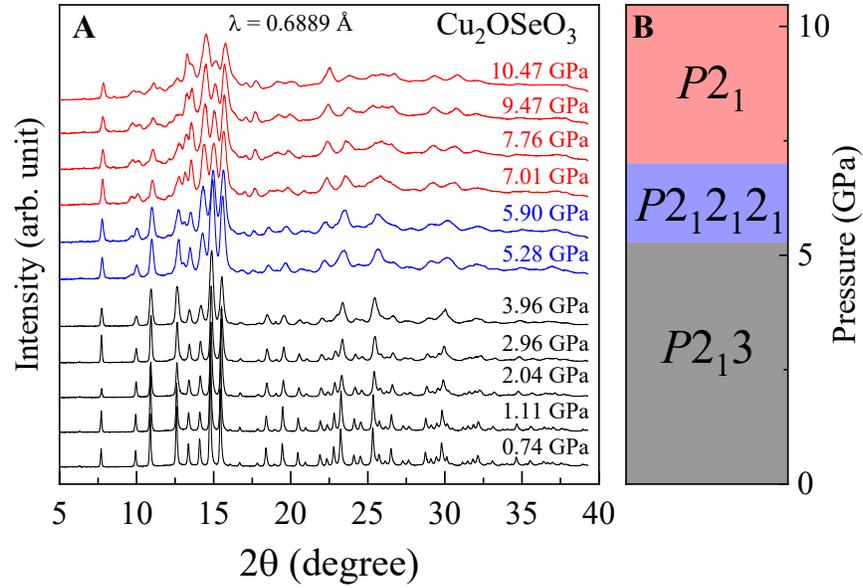

**Fig. 4. Pressure dependence of X-ray diffraction patterns.** (**A**) Evolution of room-temperature synchrotron X-ray diffraction patterns for a polycrystalline $Cu_2OSeO_3$ sample under high quasi-hydrostatic pressure up to 10.47 GPa, indicating multiple structural phase transitions. (**B**) Schematic diagram representing the pressure-induced structural phase transitions in $Cu_2OSeO_3$. It should be noted that (1) the initial cubic $P2_13$ phase transforms into the orthorhombic $P2_12_12_1$ phase at 5.28 GPa and (2) the second structural transition from the orthorhombic $P2_12_12_1$ to the monoclinic $P2_1$ occurs at 7.01 GPa.

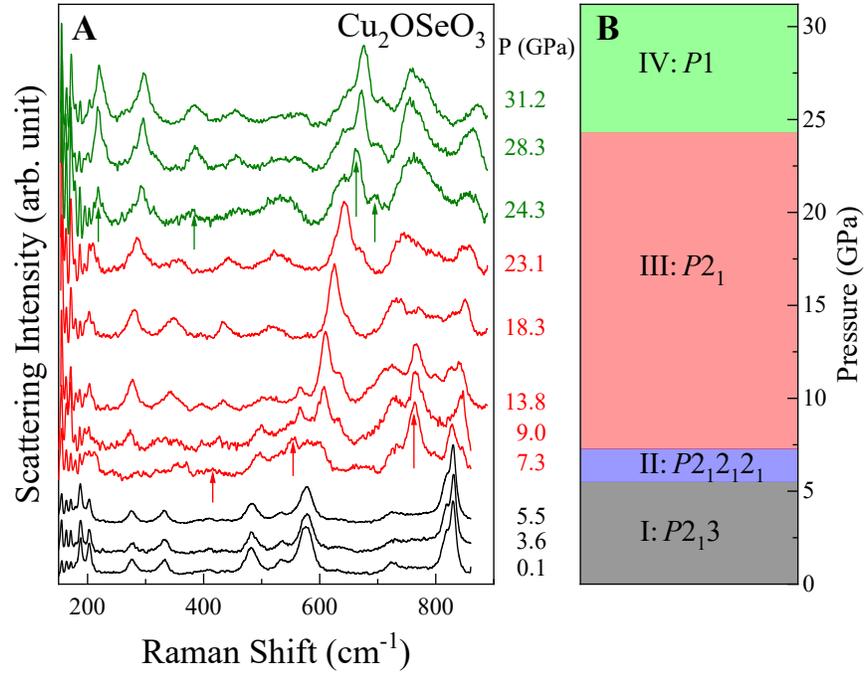

**Fig. 5. Pressure dependence of Raman spectra.** Raman spectra of $Cu_2OSeO_3$ at room temperature under different pressures up to 31.2 GPa. (**A**) Raman spectra measured with increasing pressure. (**B**) Phase diagram based on the Raman and synchrotron X-ray results. Arrows in (**A**) indicate the new frequency peaks at critical pressures.

**Supplementary Information Text**

**Section S1. Raman spectroscopy analysis under pressure.**

The structure of $Cu_2OSeO_3$ at ambient conditions is known to be cubic (space group #198, $P2_13$). The unit cell contains 56 atoms (Z=8). The Brillouin zone-center phonon modes are distributed among irreducible representations as follows: $\Gamma_{tot} = 14A + 14(^1E + {}^2E) + 42T$ (1). One of the $T$-symmetry modes is acoustic, so that $\Gamma_{Ram}$ (#198) = $14A + 14(^1E + {}^2E) + 41T$ (modes $^1E$ and $^2E$ are non-degenerate for the 23 class of representations), with all modes being Raman-active. The highest order subgroup of $P2_13$ (#198) is known to be the orthorhombic #19 ($P2_12_12_1$). It is therefore a probable candidate for the symmetry of the "new" phase at higher pressure, which is consistent with the synchrotron X-ray analysis. Considering the transition from #198 to #19, one expects the lifting of the degeneracy for $E$- and $T$-modes, so that $T \rightarrow B_1 + B_2 + B_3$, $^1E \rightarrow A$, and $^2E \rightarrow A$. Thus, $\Gamma_{Ram}$ (#19) = $42A + 41(B_1 + B_2 + B_3)$. Further, the highest order subgroup of $P2_12_12_1$ is the monoclinic $P2_1$ (#4), for which one expects the following Raman-active modes: $\Gamma_{Ram}$ (#4) = $83A + 82B$. And, finally, the subgroup of $P2_1$ is the triclinic $P1$, where $\Gamma_{Ram}$ (#1) = $165A$.

While comparing these theoretical considerations with experimental observations, one has to keep in mind that, typically, the fully symmetric (A) modes dominate the spectra because of their higher polarizability with respect to other modes. In this way one could naturally explain the appearance of "new lines" in experimental spectra of $Cu_2OSeO_3$ upon pressure-induced symmetry reduction. Indeed, as shown above, the number of A-symmetry modes increases at each consecutive step of symmetry change. The phase diagram based on Raman and synchrotron XRD results is summarized in Fig. 5B.

**Section S2. Computation of the skyrmion temperature region under pressure.**

Here we compute the critical temperatures of skyrmion phases by assuming analytical ansatz to minimize the free energy.

Our starting point is the free energy

$$f = f_0 + J(\nabla \boldsymbol{M})^2 + D\boldsymbol{M} \cdot (\nabla \times \boldsymbol{M}),$$

where the uniform part is $f_0 = a(T - T_m)M^2 + bM^4 - \boldsymbol{H} \cdot \boldsymbol{M}$ at temperature $T$ and magnetic field $\boldsymbol{H}$. As an illustration of our scheme, we first study the conical phase by the following ansatz (2)

$$\boldsymbol{M}(z) = M\{\cos(Qz) \sin \theta_0, \sin(Qz) \sin \theta_0, \cos \theta_0\},$$

where we choose $\boldsymbol{H} = \{0, 0, H\}$ to be along the $z$ axis, and the $x$ and $y$ axes are shown in Fig. 1C.

By minimizing free energy of the conical phase,

$$f[\boldsymbol{M}(z)] = a(T - T_m)M^2 + bM^4 - HM\cos\theta_0 + M^2 \sin^2\theta_0 (JQ^2 - DQ),$$

we obtain the conical angle and the wavevector of the conical phase as

$$\cos\theta_0 = \frac{H}{H_c}, Q = \frac{D}{2J}, \quad (0 < H < H_c, 0 < T < T'_m),$$

where

$$H_c = H_0\sqrt{1 - \frac{T}{T'_m}}, \quad H_0 = \frac{D^2}{2J}\sqrt{\frac{aT'_m}{2b}}, \quad \text{and } T'_m = T_m + \frac{D^2}{4Ja}.$$

The free energy density of the conical phase reads

$$f_{\text{con}} = -\frac{a^2}{4b}(T - T'_m)^2 - \frac{JH^2}{D^2}.$$

At zero field, $\theta_0 = \frac{\pi}{2}$ and the conical phase becomes the helical phase; at sufficiently large field (i.e. $H > H_c$) $\theta_0 = 0$ and the conical phase becomes the paramagnetic phase. Here, by conical ansatz, we worked out a critical temperature $T'_m$ of the conical phase, which can be written as the dimensionless form

$$T'_m = T_m + \lambda T_0, T_0 = \frac{D^2}{Ja}, \lambda = \frac{1}{4}.$$

Inspired by the results of the conical phase, we first carry out the dimensional analysis in the skyrmion phase. We find two quantities to have the dimension of temperature, $T_m$ and $T_0 = D^2/(Ja)$. As a result, we expect the critical temperatures of the skyrmion phase to be

$$T_{A1} = T_m + \xi_1 T_0, T_{A2} = T_m + \xi_2 T_0,$$

where $\xi_{1,2}$ are dimensionless parameters, which can be positive or negative, but $\xi_2 > \xi_1$ by definition. Hence, we expect the temperature range of the skyrmion phase is

$$\Delta T = T_{A2} - T_{A1} = \xi T_0,$$

where $\xi = \xi_1 - \xi_2$. Hence $\Delta T \propto T_0$, and the argument in the section "Ginzburg-Landau free energy analysis" follows.

As shown in the above calculations, the ferromagnetic phase is not favored at zero magnetic field. As shown in Fig. S3, $T'_m$ is the critical temperature of the conical phase, and $T_{A1}, T_{A2}$ for the skyrmion phase. However, due to strong fluctuations near the transition, the mean-field transition temperature $T'_m$ is replaced by the realistic and smaller value $T_c$ (3, 4). Since conical and skyrmion phases will reduce to ferromagnetism without DM interaction, their critical temperatures are determined by both the ferromagnetic transition temperature $T_m$ and the temperature scale $T_0$, which measures the strength of DM interaction, as shown in the equations above.

In the following, we will calculate this critical temperature by employing skyrmion ansatz within mean-field theory. Notice that the ansatz of the conical phase is exact since it is also a rigorous solution to the Ginzburg-Landau equations, while the ansatz of the skyrmion phase is only an approximation.

During the transition from conical to paramagnetic phases (i.e., near $T'_m$), the skyrmion phase can emerge, where line defects with opposite magnetization proliferate as shown in Fig. 1C. The upper critical temperature of the skyrmion phase can be computed by the following ansatz of skyrmions. We focus on skyrmions within each layer, where the skyrmion magnetization only depends on in-plane coordinates (2, 5). To describe the core of a skyrmion, we can employ the following ansatz written in the cylindrical coordinates $(\rho, \varphi, z)$

$$\boldsymbol{M}(\rho) = M\left\{\hat{\boldsymbol{\varphi}} \sin\left(\pi\frac{\rho}{\rho_0}\right) - \hat{\boldsymbol{z}} \cos\left(\pi\frac{\rho}{\rho_0}\right)\right\}, 0 < \rho < \rho_0.$$

The free energy of the skyrmion core compared to ferromagnetism thus reads

$$F_{\text{skm}-\text{FM}} = 2\pi \int_0^{\rho_0} \{f[\boldsymbol{M}(\rho)] - f(M\boldsymbol{z})\} \rho d\rho = \pi M^2 (12.3J - \pi D\rho_0) + \frac{\pi^2 - 4}{\pi} HM\rho_0^2.$$

We then introduce the averaged free energy density of the skyrmion core with respect to the ferromagnetic phase as

$$f_{\text{skm}} - f_{\text{FM}} = \frac{F_{\text{skm}-\text{FM}}}{\pi \rho_0^2}.$$

With the same magnitude of order parameter $M = \sqrt{\frac{a(T'_m - T)}{2b}}$, we can compare conical and skyrmion phases with respect to the ferromagnetic phase

$$f_{\text{con}} - f_{\text{FM}} = -\frac{J}{D^2}\left(H - \frac{D^2 M}{2J}\right)^2,$$

$$f_{\text{skm}} - f_{\text{FM}} = 1.76\frac{J}{D^2}H^2 - 0.59MH.$$

In fact, it can be shown that the conical phase always has lower energy than the skyrmion phase; $f_{\text{con}} < f_{\text{skm}}$ from the above expressions. However, the energy difference is very tiny: when $H = 0.29D^2M/J$, the energy difference reaches a minimum $f_{\text{skm}} - f_{\text{con}} = 0.02D^2M^2/J$. This tiny energy difference can be overcome by fluctuations, and when fluctuations increase, there is more energy in the conical phase than in the skyrmion phase. Away from the critical temperature $T'_m$, fluctuation effects are weak, and we impose the following condition for the skyrmion phase

$$f_{\text{skm}} - f_{\text{con}} \leq \Delta f_{\text{fluc}},$$

where $\Delta f_{\text{fluc}}$ is the difference in fluctuation contributions from conical and skyrmion phases to the free energy density. As a reasonable assumption, $\Delta f_{\text{fluc}}$ should be a few percent of $D^4/(J^2 b)$, the basic energy scale in the free energy expression above. When we choose $\Delta f_{\text{fluc}} = 0.01 D^4/(J^2 b)$, we can work out the lower critical temperature $T_{A1}$ of the skyrmion phase with $\xi_1 = -0.25$. The skyrmion phase is stabilized by thermal fluctuations and hence has a lower critical temperature. Close to $T'_m$, fluctuation effects are strong and can destroy all ordered phases. The exact numeric value of the higher critical temperature $T_{A2}$ is thus determined by strong fluctuations, which is beyond the scope of this work (3, 4). For simplicity, we use the mean-field results $T_{A2} = T'_m$ and $\xi_2 = 0.25$. Hence $\xi \approx 0.5$ in our ansatz. The qualitative results are shown in Fig. S3.

In the following, we apply the analysis above to different phases of the helimagnet with different structures and symmetries.

The free energy we calculated previously directly applies to Phase I

$$f_I = f_0 + D_+ W_+,$$

while at Phase II, threefold rotation symmetries are broken, which introduces another type of DM interaction (5)

$$f_{II} = f_I + D_- W_-,$$

where $W_\pm \equiv (M_x \partial_y M_z - M_z \partial_y M_x) \pm (M_z \partial_x M_y - M_y \partial_x M_z)$.

In Phase I, we have $T_0 \equiv T_I = D^2/(Ja)$, and the analysis above applies.

The additional DM interaction in Phase II makes the skyrmion anisotropic. However, the skyrmion still has two principal axes $x$ and $y$ due to the effective mirror symmetries with respect to $x$ and $y$ axes. To account for the anisotropy between $x$ and $y$ axes, we rescale our coordinates as follows

$$x' = \frac{D_{II}}{D_+ - D_-} x, \quad y' = \frac{D_{II}}{D_+ + D_-} y,$$

where $D_{II} = \sqrt{D_+^2 + D_-^2}$ and $D_+ = D$. In the rescaled coordinate system, we again consider the same ansatz of the skyrmion core written in the cylindrical coordinates $(\rho', \varphi', z)$, and the same expression of free energy $F_{\text{skm-FM}}$ compared to ferromagnetism, except that $T_0$ is now replaced by $T_{II} = D_{II}^2/(Ja) > T_I$.

At Phase III, the twofold out-of-plane rotation symmetry is broken, and there will be two additional DM interaction terms in the free energy

$$f_{III} = f_{II} + D'_+ W'_+ + D'_- W'_-,$$

where $W'_\pm \equiv (M_z \partial_x M_x - M_x \partial_x M_z) \pm (M_z \partial_y M_y - M_y \partial_y M_z)$.

The skyrmion only has twofold in-plane rotation symmetry, which maps $(x, y)$ to $(-x, -y)$. In this case, we can still find the principal axes of skyrmions as $(x'', y'')$ axes of the following rotated and rescaled coordinate system

$$\begin{pmatrix} x'' \\ y'' \end{pmatrix} = D_{III}^{-1} \begin{pmatrix} D_+ + D_- & -D'_+ + D'_- \\ D'_+ + D'_- & D_+ - D_- \end{pmatrix} \begin{pmatrix} x \\ y \end{pmatrix},$$

where $D_{III} = \sqrt{D_{II}^2 + D'^2_+ + D'^2_-}$. In the cylindrical coordinates $(\rho'', \varphi'', z)$ of the new coordinate system, we can have the same analysis of the skyrmion except that $T_0$ is now replaced by $T_{III} = D_{III}^2/(Ja) > T_{II}$.

In all of these cases, $\Delta T_i = \xi T_i$ for $i = I, II, III$, and hence we proved the results mentioned in the section "Ginzburg-Landau free energy analysis"

$$\Delta T_I < \Delta T_{II} < \Delta T_{III}.$$

At the transition to Phase IV, the skyrmion is even not invariant under in-plane twofold rotation, and analytical ansatz of the skyrmion may not be able to simplify our calculations. Nevertheless, the free energy reads (5)

$$f_{IV} = f_{III} + D''_+ W''_+ + D''_- W''_-,$$

where $W''_\pm \equiv (M_x \partial_x M_y - M_y \partial_x M_x) \pm (M_y \partial_y M_x - M_x \partial_y M_y)$. Since the total DM interactions become stronger than at the transition to Phase III, we expect $\Delta T_{IV} > \Delta T_{III}$.

The result of $T_{A1,A2}$ can be improved by better ansatz such as skyrmion lattice (6) and numerical simulations such as Monte Carlo (7). However, the qualitative behavior should not be changed significantly.

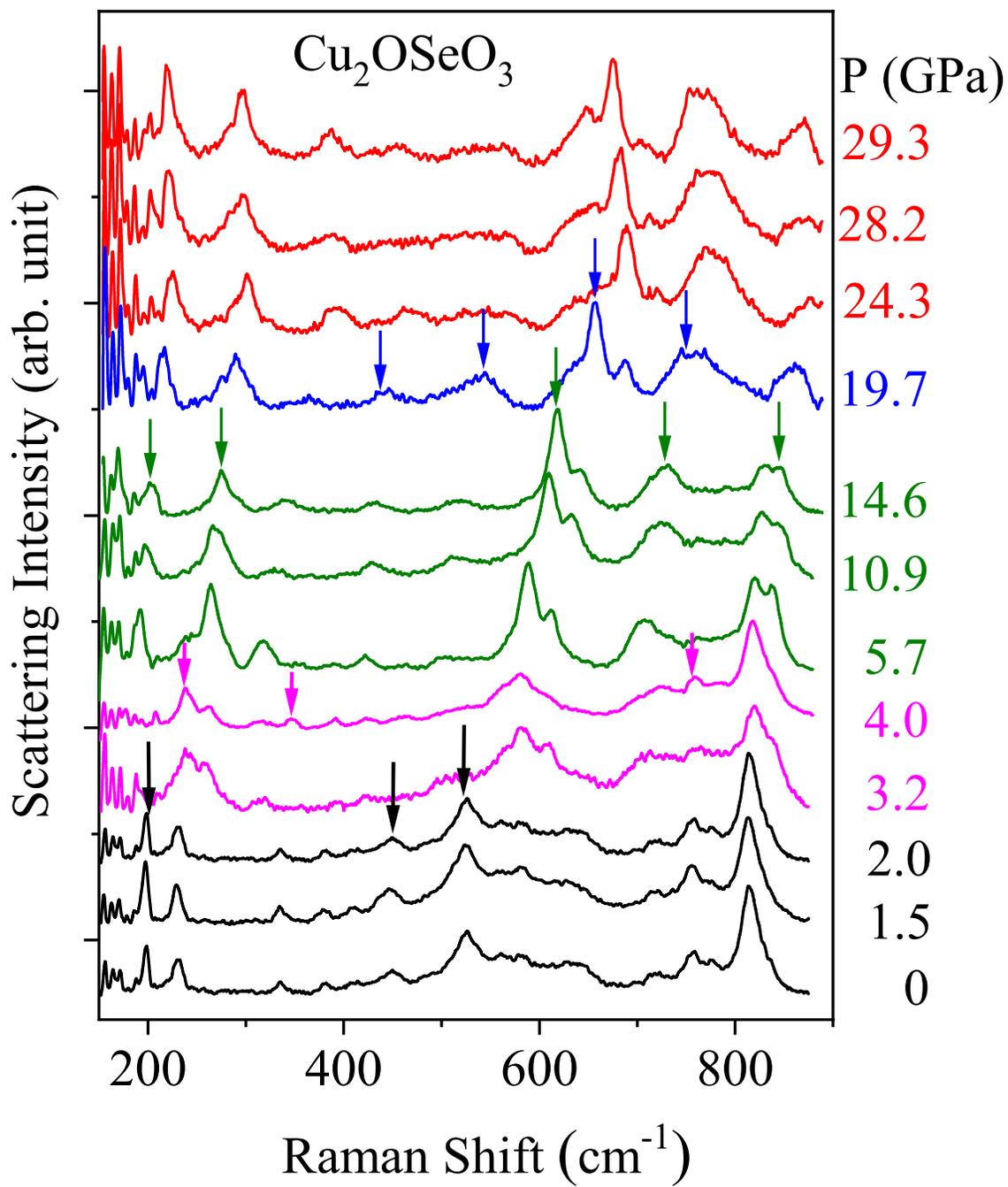

**Fig. S1.** Raman spectra under different pressures during the unloading of the pressure cell.

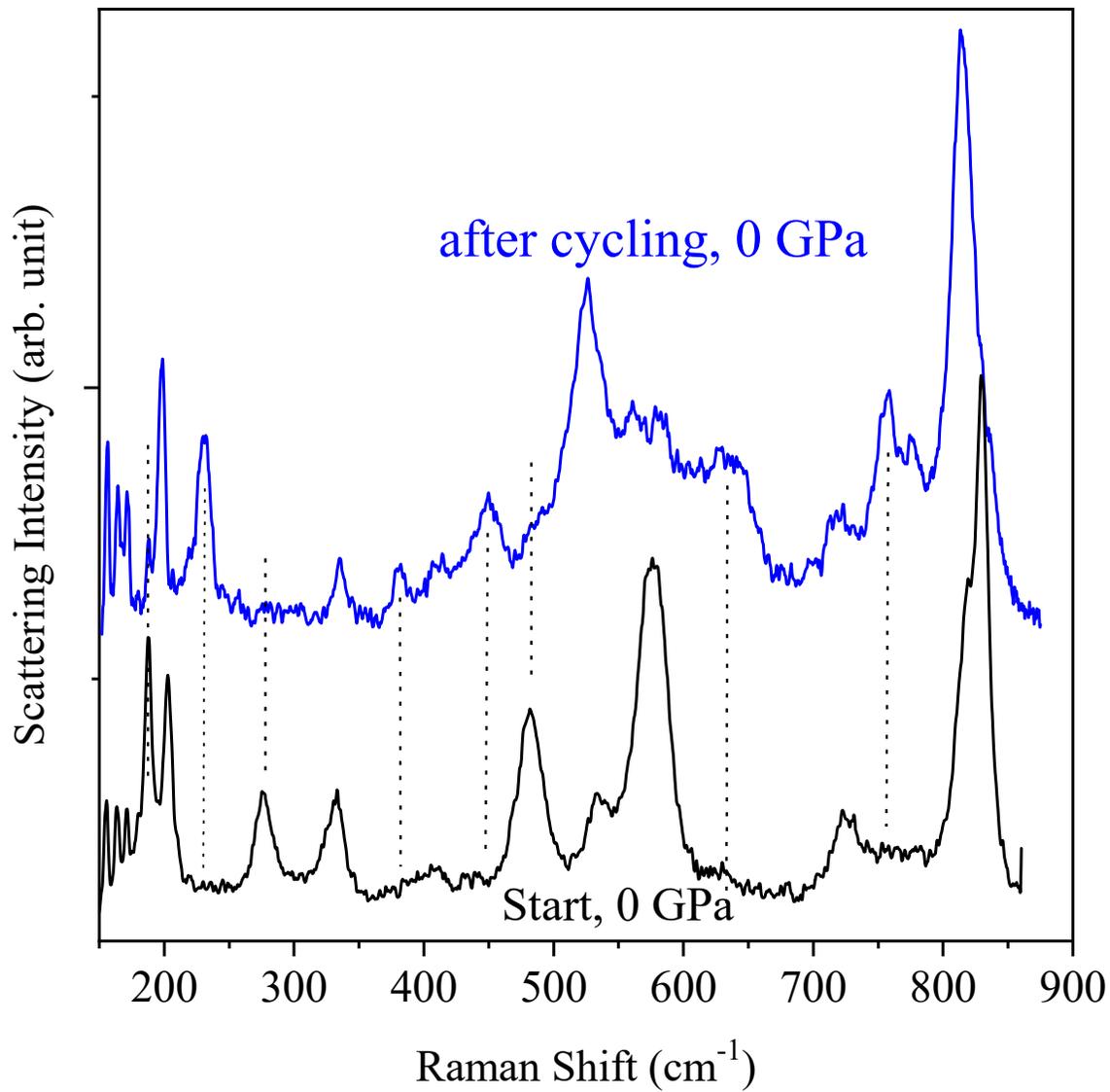

**Fig. S2. Raman spectra at ambient pressure before and after pressure cycling.** This result reveals the irreversibility of this sample after being released from high pressure.

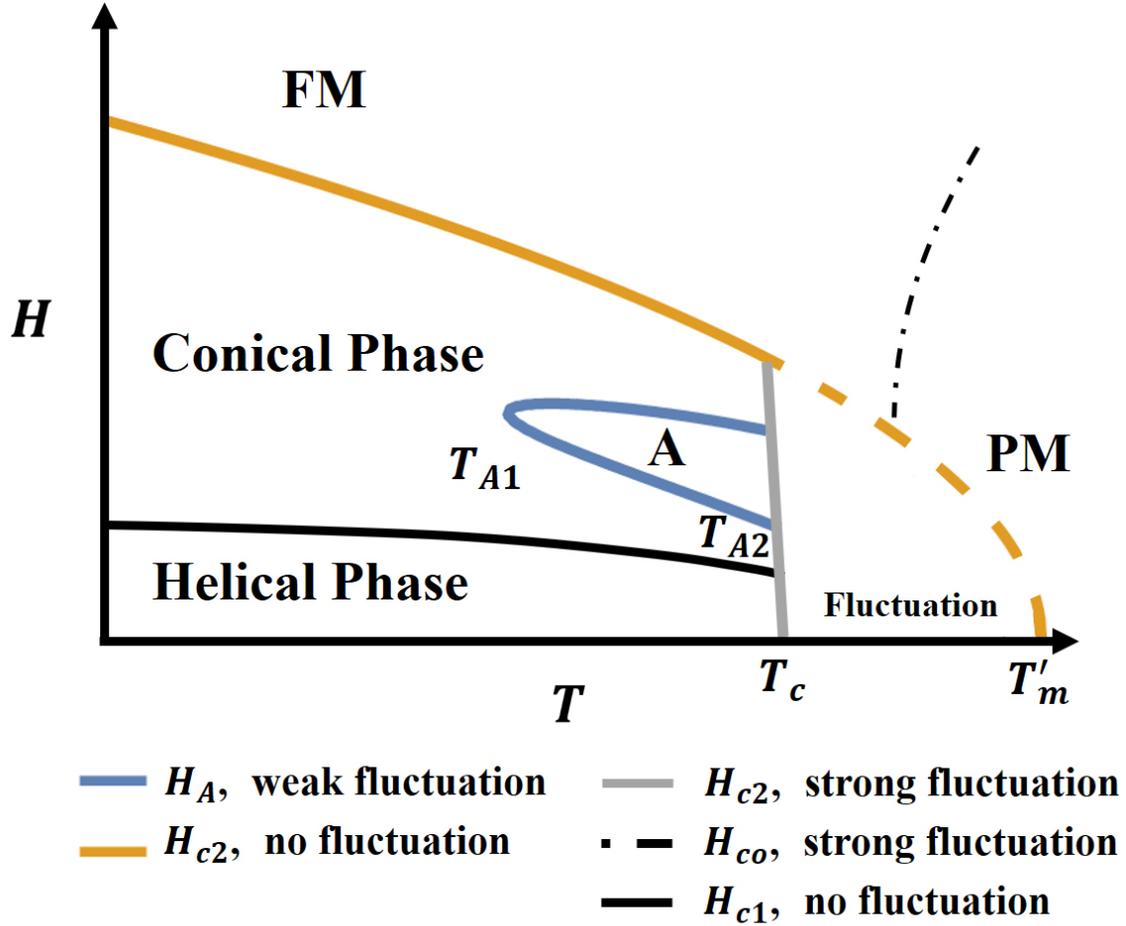

Fig. S3. Schematic phase diagram. Phase transitions and crossover are denoted by solid and dashed lines respectively. Here $T_{A1(A2)}$ is the lower (upper) critical temperature of the A phase and $T'_m$ is the mean-field transition temperature from the conical phase to the paramagnetic (PM) phase, while $T_c$ is the realistic transition temperature including the strong fluctuation effect. At high field and low temperature there is also the ferromagnetic (FM) phase. Due to strong fluctuations, near $T'_m$ there are no well-defined phase transitions but only crossovers.